\documentclass[useAMS,usenatbib]{mn2e}

\usepackage{amsmath}
\usepackage{graphicx}
\usepackage{subfig}
\usepackage{float}
\usepackage[T1]{fontenc}
\usepackage{aecompl}
\usepackage{color}
\begin{document}
\title[Roche Lobe Underfilling in PSR J102347.6+003841?]{Roche Lobe Underfilling of the Secondary Star in PSR J102347.6+003841?}
\author[O. McConnell et al.]{O. McConnell,$^1$\thanks{Email: o.mcconnell@mars.ucc.ie} P. J. Callanan,$^1$ M. Kennedy,$^{1,2}$ D. Hurley,$^1$ P. Garnavich,$^2$ \newauthor and J. Menzies$^3$.
\\$^1$Department of Physics, University College Cork, Ireland.
\\$^2$Department of Physics, University of Notre Dame, Notre Dame, IN 46556.
\\$^3$South African Astronomical Observatory, Observatory Road, Observatory, 7925, South Africa.}
\date{Accepted 2015 May 27.  Received 2015 May 26; in original form 2014 September 10}
\maketitle
\pagerange{\pageref{firstpage}--\pageref{lastpage}} \pubyear{}
\label{firstpage}

\begin{abstract}
PSR J102347.6+003841 is a radio pulsar system with a spin period of 1.69 ms and an orbital period of 4.75 hours. Uniquely, it undergoes periods of transient accretion from its companion star: it occupies an important position in the evolutionary track from X-ray binary to isolated millisecond radio pulsar.  Here we present a spectroscopic study of this system showing late-type absorption features which match those of a G2V star.  We find a semiamplitude of $286 \pm 3$ kms$^{-1}$ and a best fit orbital period of 0.1980966(1) days.  We combine these measurements with optical photometry which suggests the secondary star may be underfilling its Roche lobe by between 15\% and 20\%.  We weakly constrain the mass of the neutron star to be $\leq$ 2.2 M$_\odot$ at the 2$\sigma$ level.  We also discuss the possible origins of the H$\alpha$ emission line in our template subtracted, averaged spectrum.  Finally we present and discuss new optical photometry of J1023 taken during the recent outburst of the system.
\end{abstract}


\section{Introduction} \label{Introduction}
Low mass X-ray binaries (LMXBs) are a class of interacting binary star in which material accretes onto a neutron star or black hole from a low mass companion star.  It is believed millisecond pulsars (MSP) are formed as a result of this accretion process spinning up the neutron star.  The discovery of accreting millisecond X-ray pulsars (\citet{wijnands98}) confirmed this, however a key question is how do these systems evolve into isolated radio millisecond pulsars?
 
PSR J102347.6+003841 (hereafter J1023) is a radio pulsar with a spin period of 1.69ms and an orbital period of 4.75 hours.  It was discovered by \citet{bond02} as part of the ``Faint Images of the Radio Sky at Twenty Centimeters" (FIRST) survey.  The spectrum was dominated by a blue continuum, much like that of a Cataclysmic Variable (CV) star in quiescence, but with stronger than normal HeI and HeII lines, suggesting J1023 may be a magnetic CV.  They also realised this system lay in the region of sky covered by the Sloan Digital Sky Survey (SDSS).  They analysed the archival SDSS data and found the emission lines were double peaked, indicative of the presence of an accretion disc.  \citet{woudt04} found a repetitive modulation of V $\sim$ 0.45 mag about a mean of V $\sim$ 17.5 and an orbital period of 4.75 hours.  They also detected low amplitude flickering, suggesting on-going mass transfer.
 
J1023 was observed again by \citet{thorstensen05} (hereafter TA05) where they found only late-type absorption features, suggesting a dramatic change in the system.  They combined radial velocity estimates and optical photometry which lead them to conclude it was likely that the primary star had a mass greater than the Chandrasekhar limit and so it was actually a neutron star or black hole, rather than a white dwarf.  This hypothesis was supported by the optical and X-ray observations of \citet{homer06} of J1023, also taken while in quiescence, as they were unable to explain the observed X-ray emission in any other way.
 
In 2009 Archibald et al. detected radio pulsations from the system which confirmed the primary was indeed a neutron star and as such J1023 was confirmed as a LMXB.  They combined their binary parameters with the radial velocity measurements of TA05 to estimate the mass ratio of the system, q, to be $7.1 \pm 0.1$.  Radio eclipses were also detected, and as the line of sight between the pulsar and Earth does not intersect the Roche lobe of the companion, they concluded these must be caused by material being forced off the companion star by a pulsar wind, as seen in other similar systems (\citet{d'amico01}, \citet{hessels11}).

\begin{figure}
\includegraphics[width=84mm, angle=270, scale=0.7]{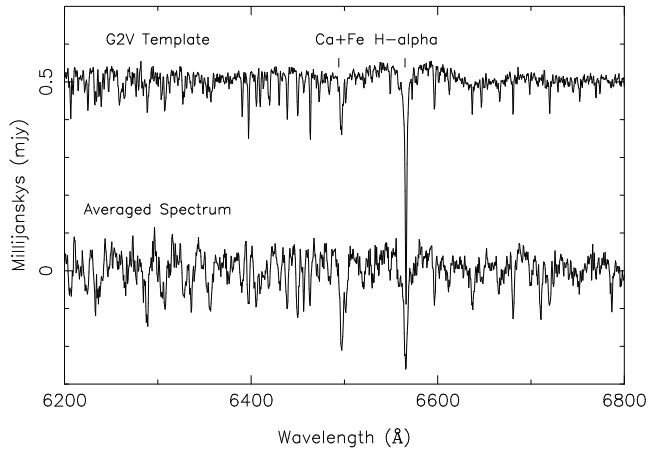}
\caption{The average of all 26 spectra is shown below and the G2V template used for cross-correlation is shown on top.  The two most prominent absorption features, the Ca+Fe blend and H-alpha lines, are labeled.  The data have been shifted to the rest frame of the secondary.}
\label{avetemp}
\end{figure}

In June 2013 J1023 became radio quiet (\citet{stappers13}), suggesting, once again, that the system had undergone a transition in state.  They saw a dramatic increase in the $\gamma$-ray flux which along with the softer X-ray spectrum of \citet{patruno14} and the reemergence of double peaked emission lines as seen by \citet{halpern13} confirms accretion, specifically via a disc, has begun again in J1023.  

The discovery of J1023 as a millisecond pulsar when in quiescence makes the system extremely interesting, especially given the double peaked emission lines seen in 2002 and from last year.  The history of the system suggests J1023 undergoes periods of transient accretion from its companion star and as such it occupies an important position on the track from LMXB to isolated MSP.  Here we present analysis of archival optical data, combined with new optical photometry, and attempt to more tightly constrain the dynamical properties in this important system.

\section{Observations and Data Reduction}

The Isaac Newton Group (ING) archival data of Rodriguez-Gil and Santander presented here were taken using the Intermediate dispersion Spectrograph and Imaging System (ISIS) optical spectrometer on the 4.2m William Herschel Telescope at La Palma, Canary Islands.  The red arm of ISIS is equipped with a 4096 x 2048 pixel array with an image scale of 0.22 arcsec pixel$^{-1}$ and the high resolution R1200R grating with a dispersion of 0.26 \AA /pixel was used.

The observations consisted of 26 600s individual exposures obtained during the night of April 12 2009 and cover the whole orbital period.  We combine this with the photometry of TA05 to better constrain the binary parameters.  Data were reduced using the standard \verb"IRAF"\footnote{IRAF is distributed by the National Optical Astronomy Observatories, which are operated by the Association of Universities for Research in Astronomy, Inc., under cooperative agreement with the National Science Foundation} routines for bias and sky subtraction, flat fielding and wavelength calibration.  The data were then exported to the \verb"MOLLY"\footnote{http://www2.warwick.ac.uk/fac/sci/physics/research/astro/ people/marsh/software} package where they were rebinned onto a common velocity scale for analysis.

We also obtained photometry of J1023 in the V-band using the 1.8m Vatican Advanced Technology Telescope (VATT) located at Mount Graham International Observatory (MGIO) on April 29 2014. The images were taken using the VATT4K CCD with a typical exposure time of 15s. A total of 329 images were taken. The images were reduced using the standard \verb"IRAF" routines for flat fielding and bias subtraction, and the relatively uncrowded field was analysed with the \verb"PHOT" command in the \verb"DIGIPHOT" package.  We also obtained three nights of V-band photometry in June 2014 using the STE3 CCD on the 1.0m telescope at the South African Astronomical Observatory (SAAO).  A total of 63 300s exposures were taken and reduced and analysed in the same way as the VATT data.  In both cases differential magnitudes were found using two stars, from TA05, in the same field of view as the pulsar: a V = 14.86 magnitude comparison star and a V = 17.40 magnitude star as a check.

\begin{table}
\centering
\begin{tabular}{c c c c}
\hline
Name     & Spectral Type & $\chi^{2}$  \\
\hline\hline
HR6538	 & G1V  & 34.023805  \\
BD+08 2015 & G2V  & 21.4970827 \\
HR1262   & G5V  & 37.3525085 \\
HR159    & G8V  & 38.4522552 \\
\hline
\end{tabular}
\caption{The range of G-type stars used in the optimal subtraction process described in the main text.  We note the minimum $\chi^{2}$ occurs for a G2V star.}
\label{temptest}
\end{table}

\section{Results}

Figure \ref{avetemp} shows the average pulsar spectrum along with our choice of template.  The spectral type of the secondary star was estimated by performing a $\chi^{2}$ fit to our selection of G-type stars.  The results are displayed in Table \ref{temptest}.  

\begin{figure}
\includegraphics[width=84mm, scale=0.75]{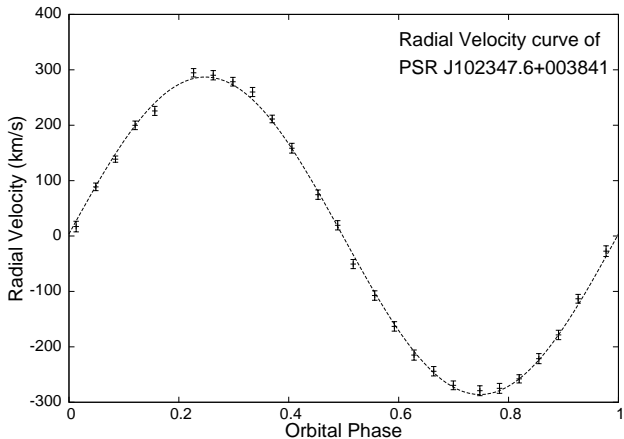}
\caption{Radial velocity curve folded over the best fit period of 0.1980966(1) days.  Error bars have been scaled to achieve a fit with a reduced $\chi^{2}$ = 1.}
\label{rvcurve}
\end{figure}

The radial velocities of the individual spectra were found through cross-correlation with a high signal-to-noise G2V template spectrum observed on the same night as the target spectra.  The wavelength range used was 6200\AA\ - 6800\AA\ (excluding the H$\alpha$ region) and the cross-correlation was carried out using the \verb"xcor" routine in \verb"molly" and \verb"fxcor" in \verb"IRAF".  Figure \ref{rvcurve} shows the radial velocity curve obtained through the cross-correlation technique.  A sine wave fit yields a systemic velocity of $\gamma$ = $0.45 \pm 1.6$ kms$^{-1}$, in good agreement with TA05, and a semi-amplitude of $286 \pm 3 $ kms$^{-1}$ for K$_{2}$ compared to $268 \pm 4 $ kms$^{-1}$ found by TA05.  However our value of K$_{2}$ supersedes theirs as our spectra are of a higher resolution (1\AA\ compared to their 3.5\AA) and due to the better sampling of our radial velocity curve.

In order to refine the orbital period of TA05 we used the O-C method described by \citet{bevington}.  This resulted in a best fit orbital period P = 0.1980966(1) days at the 1$\sigma$ confidence limit.  This improves on the value found by TA05, however not on the value found by \citet{archibald09} of 0.1980962019(6) days which was obtained through timing measurements of radio pulsations of the pulsar.

Using our K$_{2}$ and orbital period we estimate the mass function, f(m$_2$), to be $0.48 \pm 0.02$ M$_\odot$.

\begin{table}
\centering
\begin{tabular}{c c}
\hline
Free Parameters	& Results\\
\hline\hline
$f$		& $0.83 \substack{+0.03\\-0.02}$\\
${i}$   	& $54 \degr \substack{+1 \degr\\-5 \degr}$\\
T$_{eff}$   	& $5655 \pm 85$ K\\
L$_{x}$		& $9.9 \substack{+0.4 \\ -1.6} \times$ 10$^{33}$ erg s$^{-1}$\\
\hline
\end{tabular}
\caption{Results of the ELC modelling for the secondary star underfilling its Roche Lobe.}
\label{elcresults}
\end{table}

\section{An Eclipsing Light Curve Model}

\begin{figure}
\includegraphics[width=84mm, scale=0.75]{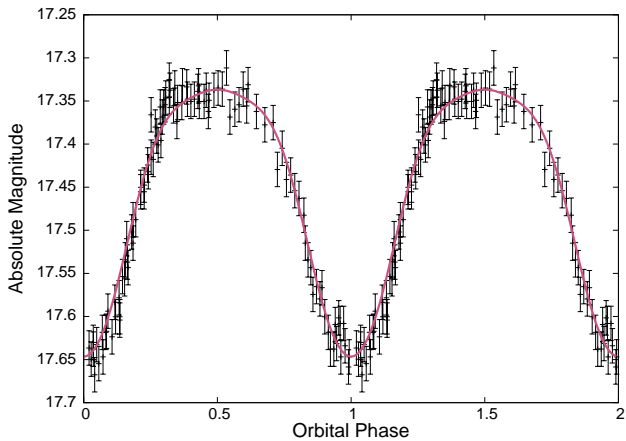}
\caption{Results of our Roche lobe underfilling modelling of J1023 using ELC and the V-band photometry of TA05.  Data are repeated for clarity.}
\label{thorsphot}
\end{figure}

We next attempted to constrain the system parameters by utilising the Eclipsing Light Curve Code, (ELC), of \citet{orosz}, to simultaneously fit the V-band light curve of TA05 and our new radial velocity curve. A Monte Carlo Markov Chain (\citet{tegmark14}) was used to determine errors.  The bounded free parameters were the Roche lobe filling factor, f, the effective temperature of the secondary, 5570K $< T_{eff} <$ 5750K (in line with a G2V spectral type), the irradiating luminosity of the pulsar, 10$^{33}$ - 4 $\times10^{34}$erg s$^{-1}$ (where the upper limit is determined by the spin down luminosity of the pulsar), the mass ratio of the system, $7.0 < q < 7.2$ (we adopted the mass ratio of \citet{archibald09}, $7.1 \pm 0.1$), and the inclination of the system, $36\degr < i < 55\degr$. The upper bound on the inclination comes from previous estimates of \citet{archibald09} while the lower bound comes from noting that, for a fixed K$_2$, period and mass ratio, the calculated mass of the primary is entirely dependent on the inclination of the system, and an inclination of less than 36$\degr$ gives a primary mass greater than 3 M$_\odot$, the maximum theoretical mass of a neutron star (\citet{chamel13}).  The albedo of the secondary star was set to 0.5, in keeping with the star being a low mass (e.g. $\sim$0.2 M$_\odot$ for a 1.4 M$_\odot$ primary), convection dominated star as noted by \citet{deller12}.

\begin{figure*}
\begin{minipage}{168mm}
\includegraphics[width=168mm]{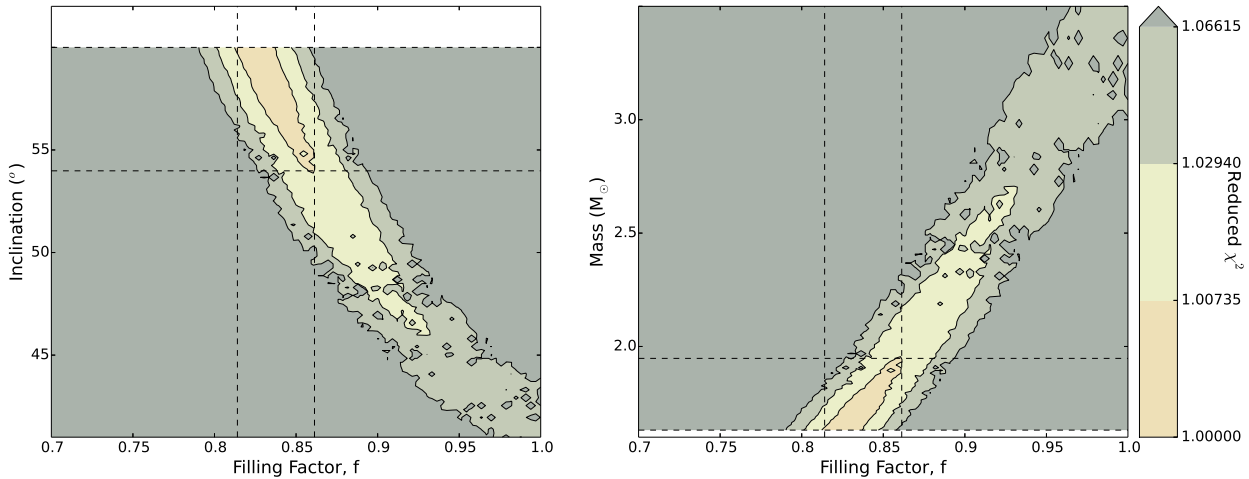}
\caption{The left panel shows the inclination of the system against the Roche lobe filling factor of the secondary star and the right panel shows the mass of the primary against the filling factor.  The contours here represent constant values of reduced $\chi^{2}$ and the dashed lines are the 1$\sigma$ confidence limits.  The $\chi^{2}$ minimum occurs at 135.88 which corresponds to a reduced $\chi^{2}$=1.}
\label{contours}
\end{minipage}
\end{figure*}

The results can be seen in Table \ref{elcresults}, with the fit to the V-band shown in Figure \ref{thorsphot}.  The left panel in Figure \ref{contours} shows the correlation between the inclination of the system and the filling factor of the companion. The inclination here is higher than the $44 \degr \pm 2 \degr$ inferred from \citet{deller12} however they have a higher mass for the neutron star which assumed a Roche lobe filling secondary. As we have allowed the Roche lobe filling factor of the secondary star to vary, we are only able to provide an upper limit to the mass of the neutron star. The limits are 1.4 M$_\odot$ at the 1$\sigma$ level and 2.2 M$_\odot$ at the 2$\sigma$ level.  This can be seen in the right panel in Figure \ref{contours} which shows the correlation between the mass of the neutron star and the filling factor of the companion.

It is worth noting that our modelling assumes the secondary star has a convective envelope which, given the evolutionary history of the system, may not necessarily be the case.  We note that a radiative envelope changes the optical modulation by approximately 0.1 magnitudes which would have a significant effect on the parameters estimated here.

We also investigated the effects of veiling on the light curve (for example, due to a constant contribution from an accretion disk).  For a Roche lobe filling factor of one we require a relatively high veiling (of order 100$\%$, or twice the observed minimum flux) to best match the observed data. However, this is not consistent with veiling constraints we obtain from spectroscopic measurements: these provide an upper limit of 30$\%$ for a G2V template.

By using a temperature of 5655K for a G2V star and radius of the companion star as computed by ELC of $0.39 \pm 0.07$ R$_\odot$, we found the absolute magnitude of the companion should be $6.88 \pm 0.07$, which is in good agreement with the absolute magnitude of $6.84 \pm 0.07$, calculated using a distance to the system of $1368\substack{+42\\-39}$ pc from \citet{deller12} and using an extinction of 0.14 in the V band. 

\subsection{Residual H$\alpha$ Emission}

\begin{figure}
\includegraphics[width=80mm, angle=270, scale=0.75]{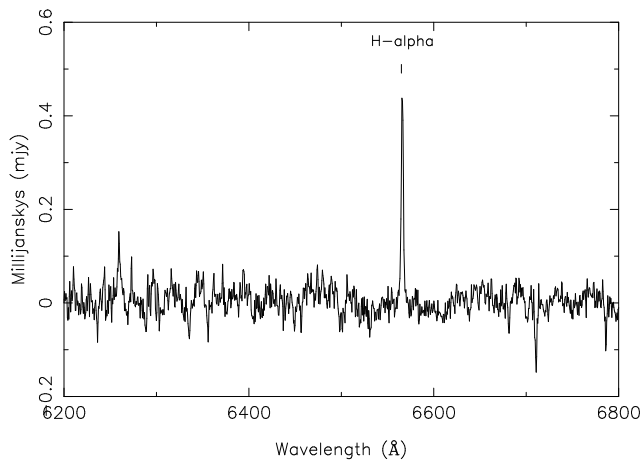}
\caption{The residual H$\alpha$ emission seen as a result of the optimal subtraction.}
\label{residha}
\end{figure}

The cross correlation, averaging of the J1023 spectra and optimal subtraction of the template spectrum resulted in a residual H$\alpha$ emission line with equivalent width $(-1.3 \pm 0.2)$ \AA (where the error was found by varying the continuum level when fitting the emission line) and can be seen in Figure \ref{residha}.  To estimate the K$_2$ of the H$\alpha$ emission line we cross correlated and optimally subtracted the template spectrum from each of our 26 individual exposures to produce a radial velocity curve of the emission line.  We found a K$_2$ of $290 \pm 4$ kms$^{-1}$ which is consistent with the absorption line value ($286 \pm 3$ kms$^{-1}$), which suggests that the source of the residual emission is from the secondary itself, possibly an irradiated inner hemisphere.  Alternatively as discussed in \citet{hernandez10}, and the references therein, the measured equivalent width is consistent with H$\alpha$ emission due to a chromospherically active secondary star.

\section{A Change in State}

\begin{figure}
\subfloat{\includegraphics[width=80mm, scale=0.7]{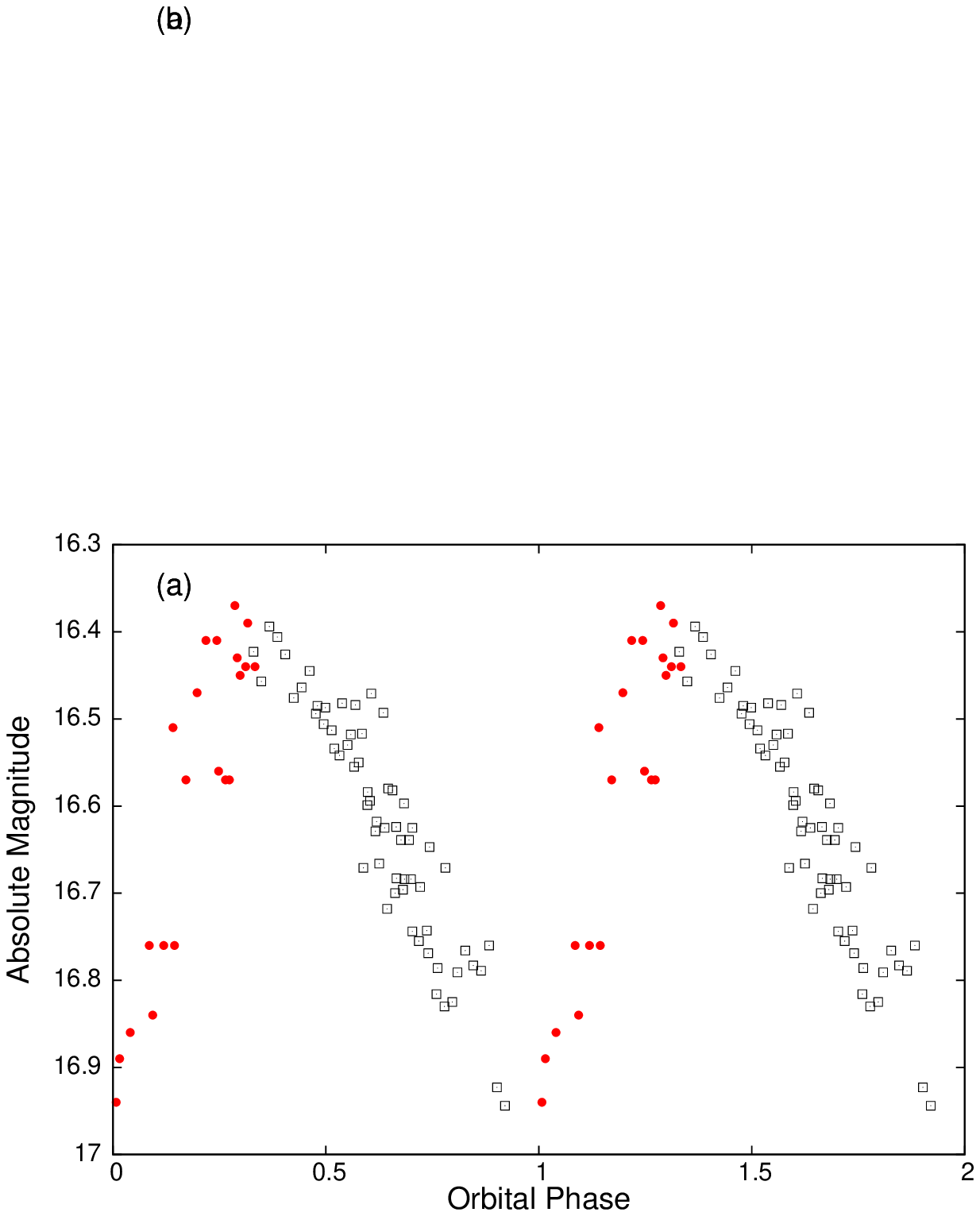}\label{saaophot}}\\
\subfloat{\includegraphics[width=80mm, scale=0.7]{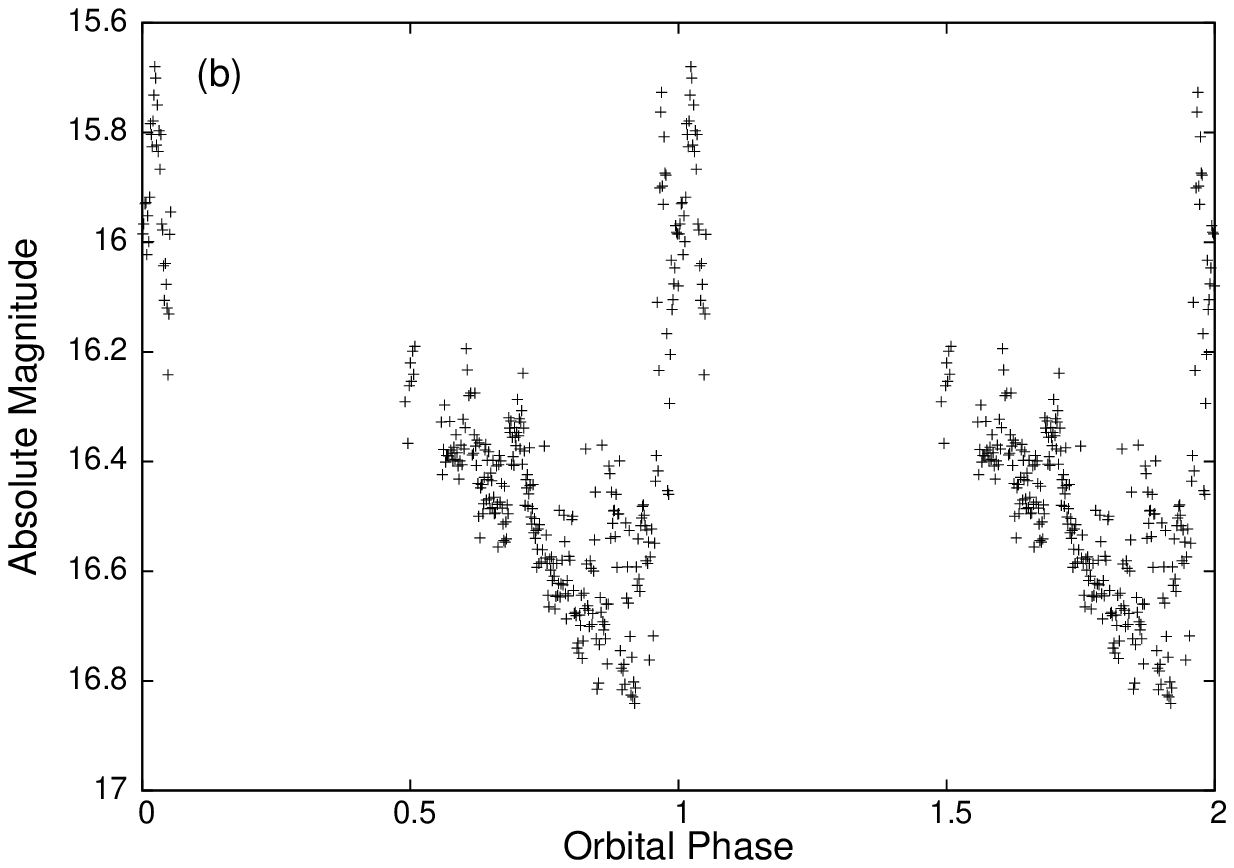}\label{vattphot}}\\
\caption{The three sets of V-band photometry taken while the system was in its recent outburst phase.  Figure \ref{saaophot} shows the SAAO data (black open squares) taken using the 1m telescope on the nights of 2014/06/06, 2014/06/12 and 2014/06/13 and Atel data (red filled circles) of \citet{halpern13} taken using the 1.3m McGraw-Hill telescope at the MDM Observatory on Kitt Peak.  Figure \ref{vattphot} shows the VATT data taken on the night of 2014/04/29.  Data are repeated for clarity.}
\label{outburstphot}
\end{figure}

As stated in Section \ref{Introduction}, J1023 reverted back to its accretion phase in June 2013.  We obtained optical photometry of the system over three nights using the 1m telescope at the South Africa Astronomical Observatory (SAAO).  This was combined with a single night's observation at the Vatican Advanced Technology Telescope (VATT) at the Mount Graham International Observatory (MGIO) in Arizona.  Together these observations cover three quarters of the orbital period and we combine this with the V-band data of \citet{halpern13} which gives us complete phase coverage (Figure \ref{outburstphot}.)
 
The data indicate that the system has brightened by over a magnitude compared to the average V-band magnitude of $\sim$17.5 when in quiescence. 

The data suggest two types of behaviour in this state: one appears as a somewhat smooth, albeit asymmetric light curve, with minimum at phase zero, whereas during the epoch of the VATT observations, a second type of more rapid, flaring variability is observed.

\subsection{Comparison to Similar Systems}

\subsubsection{Long Term Variability}

XSS J12270-4859 (hereafter J12270) was discovered as part of the Rossi X-ray Timing Explorer Project (RXTE) slew survey (\citet{sazonov13}) and was initially classified as a CV by \citet{masetti06} due to the presence of optical emission lines. Further multi-wavelength observations cast doubt on this, indeed \citet{martino10} noticed the position coincided with a Fermi $\gamma$-ray source (1FGL J1227.9-4852) which lead \citet{hill11}, amongst others, to suggest that J12270 may harbour an active radio MSP.  In their paper \citet{bassa14} describe changes in the system as it transitions from an accreting LMXB phase to that of a rotation powered MSP.  They note that the optical light curve decreases by between 1.5 and 2 magnitudes while there is a corresponding decrease, by a factor of 10, in the X-ray count rate and a decrease in the $\gamma$-ray brightness, again by a factor of between 1.5 and 2.  This matches J1023 very well as in its change of state from MSP to LMXB the optical luminosity increases by between 1 and 2 magnitudes from quiescence, the X-ray luminosity increases from $\sim$10$^{32}$ erg s$^{-1}$ to $\sim$10$^{33}$ erg s$^{-1}$ while the $\gamma$-ray flux increased by a factor of five.

\subsubsection{Short Term Variability}

Figure \ref{vattphot} shows that some of the flaring events occur on timescales shorter than the sampling time of 15s, while other, higher amplitude, flares last $\sim$ 100s.  While in similar systems both the short and long timescale variations tend to last longer than in J1023, \citet{hynes02}, in their study of V404 Cyg, suggest that shorter timescale flickering may be a common feature of LMXB systems while the larger amplitude events are more difficult to explain, but may be due to instabilities in the accretion disc.  Similar short and long timescale events are seen in Cen X-4 and \citet{shahbaz10} compared the observed colours to those predicted by three emission mechanisms: blackbody, synchrotron radiation and an optically thin layer of Hydrogen.  They favour this third mechanism and attribute the source of the flares to the accretion disc.  However the amplitude of the flaring we observe in J1023 appears unusual in comparison to other black hole or neutron star related systems.

\citet{archibald10} found evidence of X-ray emission modulated at the rotational period of the pulsar. \citet{russell07} analysed optical/infrared (OIR) and X-ray data from 19 Neutron Star X-ray Binaries (NSXB) to compare the observed OIR and X-ray fluxes to their expected values if the OIR emission is dominated by thermal emission from a X-ray heated disc or synchrotron emission from the jet.  They found that thermal emission due to X-ray reprocessing could explain all their data except at high luminosities.  Synchrotron emission from the jet was found to dominate the NIR light curve above L$_x$ $\sim$ 10$^{36}$ erg s$^{-1}$ and the optical above L$_x$ $\sim$ 10$^{37}$ erg s$^{-1}$ . This suggests that the flaring seen in our VATT data is not due to synchrotron emission, given the spin down luminosity of the pulsar, L$_x$ $\sim$ 10$^{34}$ erg s$^{-1}$.

\section{Conclusions}

The results of our modelling suggest that the secondary star in J1023 is underfilling its Roche lobe between 15\% and 20\%, albeit at the 1$\sigma$ level.  If confirmed, this suggests that the mode of accretion in this system is more likely to be via a wind from the secondary star, rather than Roche lobe overflow which is more commonly the case.  This may in turn be related to the accreting/non-accreting transitions that have been observed from this system.

We also observe evidence for H$\alpha$ in emission in the quiescent state, which we associate with either the irradiation of the secondary star by the spin down luminosity of the pulsar, or from an active chromosphere on the secondary.

As our photometric observations of J1023 in this outburst phase have been taken over several epochs, in which only part of the orbital phase has been sampled, more systemic, multi-wavelength, observations are required.  Further multi wavelength observations once it returns to quiescence are encouraged, especially phase-resolved spectroscopy and more accurate photometry to confirm the underfilling of the Roche Lobe of the secondary star, the source of the H$\alpha$ emission, and more tightly constrain the mass of the neutron star. 
 
\section*{Acknowledgments}
OM and PJC acknowledge financial support from Science Foundation Ireland.  MK and PJC acknowledge financial support from the Naughton Foundation.  We thank the Vatican Observatory and Richard Boyle for providing us with observing time on the VATT, the PLANET collaboration for providing the time for the SAAO observations and J. Thorstensen for providing us with his optical photometry.  Finally we acknowledge the use of the \verb"MOLLY" software package developed by T. Marsh.

\label{lastpage}
\end{document}